\documentclass[letter]{jpsj3}
\usepackage{txfonts,tabularx,color}

\title{Effects of Annealing under Tellurium Vapor for Fe$_{1.03}$Te$_{0.8}$Se$_{0.2}$ Single Crystals}

\author{Yuta KOSHIKA$^1$\thanks{E-mail: y.koshika1214@gmail.com}, Tomohiro USUI$^1$, Shintaro ADACHI$^1$, Takao WATANABE$^1$, Kouhei SAKANO$^2$, Shalamujiang SIMAYI$^2$, and Masahito YOSHIZAWA$^2$}
\inst{$^1$Graduate School of Science and Technology, Hirosaki University, 3 Bunkyo, Hirosaki, 036-8561 Japan \\
$^2$Graduate School of Engineering, Iwate University, Morioka 020-8551, Japan \\
} 

\abst{In this study, the effects of annealing under tellurium vapor have been investigated for Fe$_{1.03}$Te$_{0.8}$Se$_{0.2}$ single crystals. The as-grown crystal is not superconducting. After annealing it at 400 $^\circ$C for 250 h in tellurium (Te) vapor, it shows metallic resistivity temperature dependence below 150 K and a sharp superconducting transition at 13 K. The sample also shows a full magnetic shielding effect at low temperatures. The improvement in superconductivity has been attributed to the removal of excess Fe, which is inevitably incorporated in the as-grown crystals, by the annealing.}

\kword{Fe-based superconductor, Fe$_{1+y}$Te$_{x}$Se$_{1-x}$ single crystal, annealing effect, resistivity, magnetic susceptibility, electron probe microanalysis (EPMA)}

\begin{document}
\maketitle


Fe-based superconductors have attracted much attention, since their superconducting transition temperature $T_{c}$ has been raised up to 55 K\cite{ren}, which is second only to high-$T_{c}$ cuprates, and they contain magnetic Fe element, which is usually considered to be harmful for superconductivity. Among the many Fe-based superconductors, Fe$_{1+y}$Te$_{x}$Se$_{1-x}$ system\cite{fang} is suitable for basic research, since the crystal structure is simplest, that is, it consists of only conducting Fe\textit{X} (\textit{X}:Te or Se) layers. It is also less toxic compared with other FeAs-based systems, although its $T_{c}$ is relatively lower ($T_{c}$ = 14 K). Large and high quality single crystals are needed for the investigation of the superconducting mechanism. Some amount of excess Fe (represented as $y$ in the above formula), however, is inevitably incorporated in the crystals\cite{fang}. It partially occupies the so-called Fe(2) site between neighboring Fe\textit{X} layers\cite{bao} and is thought to degrade the superconducting properties of Fe$_{1+y}$Te$_{x}$Se$_{1-x}$\cite{liu}. 

Sales {\it et al.} first reported growth of Fe$_{1+y}$Te$_{x}$Se$_{1-x}$ single crystals with $0<y<0.15$ for $x\geq0.5$ using the Bridgman method\cite{sales}. They have shown that only crystals grown with a composition near $x=0.5$ exhibit bulk superconductivity, although resistivity measurements show traces of superconductivity near 14 K for all $x$ except $x=1$. It has been found that the $y$ values monotonically decrease from $y = 0.13$ for $x = 1$ to $y = 0$ for $x = 0.5$\cite{sales}. Therefore, it has been pointed out that the appearance of bulk superconductivity correlates with smaller values of $y$\cite{sales}. Soon after their study was reported, single crystals with a wide range of $x = 0.5-0.9$ were found to show bulk superconductivity when they were annealed in vacuum (hereafter, called "vacuum-anneal") at 400 $^\circ$C for more than 100 h\cite{noji1,noji2,taen}. In this case, vacuum-annealing does not result in a decrease of Fe content. It is considered that vacuum-annealing makes the distribution of Se and Te more homogeneous\cite{taen} and relaxes lattice distortion\cite{noji2}, causing bulk superconductivity. However, we could not reproduce the same positive results as Noji {\it et al.}\cite{noji1}, although some improvement in superconductivity was observed as described below. This suggests that the effect of vacuum-annealing depends on the quality of the as-grown crystals and/or detailed annealing conditions. Recently, on the other hand, it has been reported\cite{dong,kawasaki} that the superconductivity is remarkably improved when the crystals are annealed in air or oxygen, especially when there is a lower Se concentration. The energy-dispersive x-ray (EDXS) analysis has shown that the Fe content in the crystals annealed in air is a little less than that in the as-grown crystals or those annealed only in a vacuum, indicating that the excess Fe atoms are partially removed by the annealing\cite{dong}. The reduction of excess Fe is not complete, and in addition, some specific annealing conditions are required to obtain good results\cite{dong}. Long time annealing causes degradation of the superconducting properties of the crystals\cite{kawasaki}.

In this paper, we report on the effects of annealing under tellurium (Te) vapor for Fe$_{1.03}$Te$_{0.8}$Se$_{0.2}$ single crystals. The non-superconducting as-grown crystals exhibit a full shielding effect after they have been annealed. Our chemical composition analysis using an electron probe microanalyzer (EPMA) indicates that the excess Fe in the crystal is completely removed.  

\begin{figure}
\includegraphics[width=80mm]{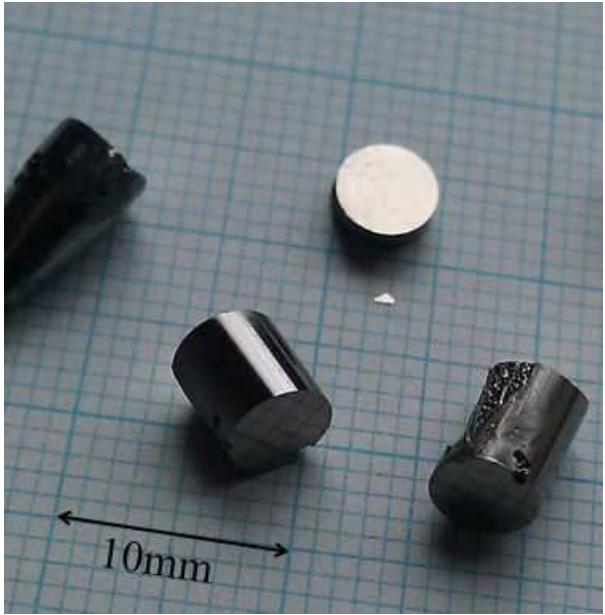}
\caption{\label{f1} (Color online) Photograph of as-grown Fe$_{1.03}$Te$_{0.8}$Se$_{0.2}$ single crystals. The composition is a nominal value.}
\end{figure}

Single crystals with nominal composition Fe$_{1.03}$Te$_{0.8}$Se$_{0.2}$ were grown using the Bridgman method. Initial materials were Fe (99.9$\%$) powder, Te shot (99.0$\%$), and Se (99.9$\%$) powder. An appropriate amount was mixed thoroughly in an glove box. The mixed powders were loaded into a small quartz tube with $d = 6$ mm$\phi$, and the tube was sealed into another large-sized evacuated quartz tube. The quartz ampoule was heated at 1050 $^\circ$C for 85 h, and then cooled down to 550 $^\circ$C at the rate of 4 $^\circ$C/h under a temperature gradient of around 8$^\circ$C/cm, and finally furnace cooled down to room temperature. Large-sized (6 mm$\phi$ x 20 mm) single crystals were obtained as shown in Fig. \ref{f1}. 

In order to anneal the as-grown crystals under Te vapor (hereafter, called "Te-anneal"), they were sealed with pulverized Te, while crystals and pulverized Te were detached from each other, into a small pyrex tube under a vacuum. The pyrex tube was heated at 400 $^\circ$C for 150 h and 250 h. Sufficient amounts of Te were loaded in order to fill the space in the tube at its equilibrium vapor pressure. Some type of Te-compound was observed to adhere to the annealed crystals making them grayish. Those compounds were cleaved away before characterization of the crystals. The surface of the cleaved crystals was always flat and shiny. Vacuum-annealing was also carried out for comparison.

The as-grown and annealed crystals were characterized by X-ray diffraction (XRD), resistivity, magnetic susceptibility, and EPMA measurements. Single crystal XRD measurements were performed using a diffractometer with Cu-K$_\alpha$ radiation. The resistivity measurements were carried out by the DC four-terminal method. The magnetic susceptibility measurements were performed using a superconducting quantum interference device (SQUID) magnetometer (Quantum Design MPMS) with the magnetic field applied parallel to the thin sample (perpendicular to the c-axis). The chemical composition was determined by the EPMA measurements. In order to increase the accuracy of the measurements, polycrystalline FeTe$_{0.6}$Se$_{0.4}$ (nominal) was synthesized beforehand by quenching a high temperature solution, and that was used as a standard sample. The composition was determined by averaging data of 5-6 regions with a spot size of 10 $\mu$m.   

\begin{figure}
\includegraphics[width=80mm]{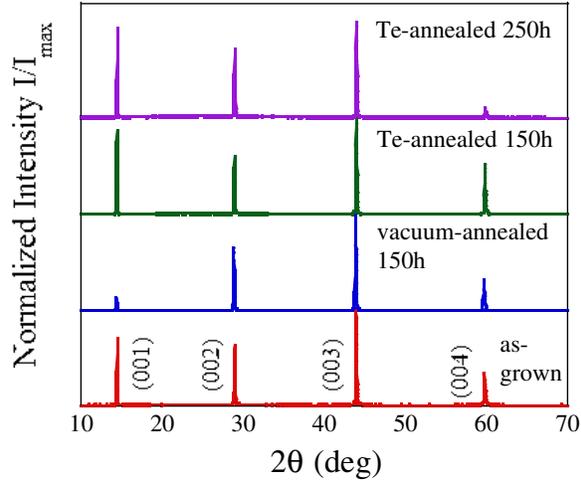}
\caption{\label{f2} (Color online) X-ray diffraction patterns of as-grown and annealed Fe$_{1.03}$Te$_{0.8}$Se$_{0.2}$ (nominal composition) single crystals.}
\end{figure}
\begin{figure}
\includegraphics[width=80mm]{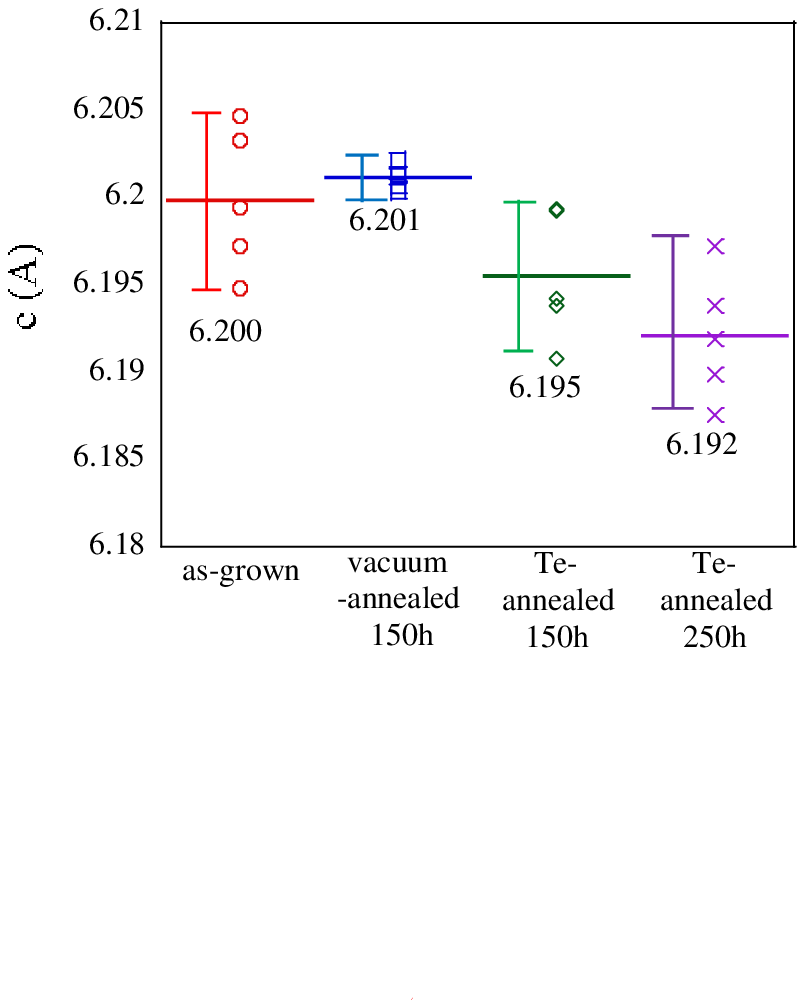}
\caption{\label{f3} (Color online) c-axis lengths of as-grown and annealed Fe$_{1.03}$Te$_{0.8}$Se$_{0.2}$ (nominal composition) single crystals.}
\end{figure}

Figure \ref{f2} shows XRD patterns for the as-grown and annealed Fe$_{1.03}$Te$_{0.8}$Se$_{0.2}$ (nominal composition) single crystals. All peaks are assigned to $(00l)$ peaks of Fe$_{1+y}$Te$_{x}$Se$_{1-x}$. These results indicate that each crystal is a phase pure Fe$_{1+y}$Te$_{x}$Se$_{1-x}$ single crystal. The c-axis length is estimated by using a fitting method of the Nelson-Riley (N-R) function. All $(00l)$ peaks of each crystal are well fitted into one line indicating that their crystallinity is good. Here, we have measured five samples of each type. All the data and the averaged values are shown in Fig. \ref{f3}. These averaged values agree with those of Fe$_{1+y}$Te$_{x}$Se$_{1-x}$ for $x \approx 0.8$\cite{fang,kawasaki}. A slight tendency for the c-axis length to shrink is observed by the Te-annealing. This Te-annealing effect on XRD will be discussed below. 
 
\begin{figure}
\includegraphics[width=80mm]{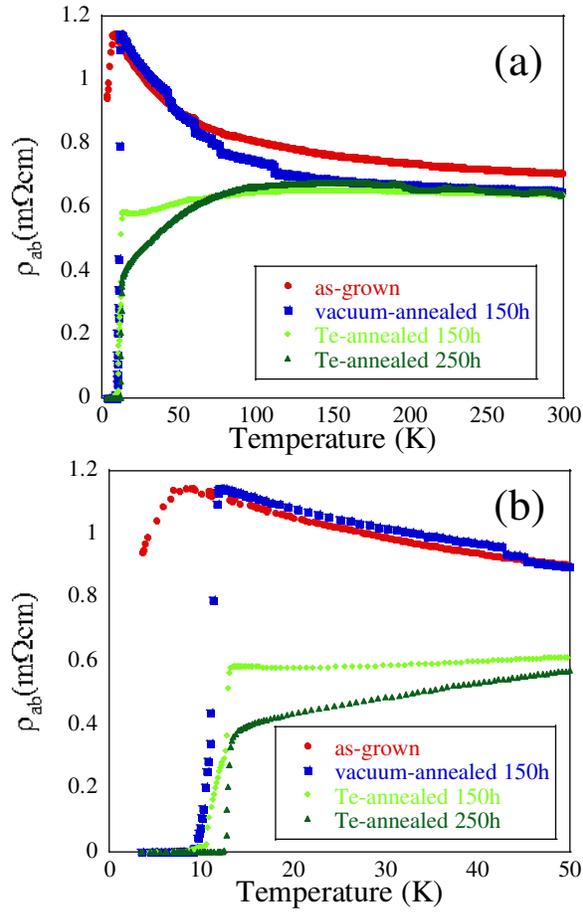}
\caption{\label{f4} (Color online) (a) Temperature dependence of the in-plane resistivity, $\rho_{ab}$, of as-grown and annealed Fe$_{1.03}$Te$_{0.8}$Se$_{0.2}$ (nominal composition) single crystals. (b) An enlarged plot of $\rho_{ab}$ near $T_{c}$.}
\end{figure}

Figure \ref{f4} (a) shows the temperature dependence of the in-plane resistivity, $\rho_{ab}$, of as-grown and annealed Fe$_{1.03}$Te$_{0.8}$Se$_{0.2}$ (nominal composition) single crystals. Figure \ref{f4} (b) shows the enlarged plot near $T_{c}$. The as-grown sample shows semiconductor-like behavior for almost all temperature regions measured, while it starts to drop below 6 K but does not reach zero resistivity. The vacuum-annealed sample shows zero resistivity at 10 K, while the normal state behavior is similar to that of the as-grown sample. As for $\rho_{ab}$ in the sample annealed in the Te vapor for 150 h, the semiconductor-like behavior is suppressed and zero resistivity appears at 11 K. However, the transition is two-phase-like (Fig. \ref{f4} (b)), suggesting the sample is inhomogeneous due to insufficient annealing. Therefore, the annealing is prolonged to 250 h. Then, $\rho_{ab}$ shows metallic temperature dependence below 150 K and superconductivity is observed at 13 K. This metallic behavior is very similar to that of a more Se-rich sample, FeTe$_{0.6}$Se$_{0.4}$, which has very little or no excess Fe atoms, as reported by many authors\cite{taen,noji1,dong,lei,khim}. To the best of our knowledge, this result is the first example of bulk Fe$_{1+y}$Te$_{x}$Se$_{1-x}$ single crystals with a lower Se concentration ($x \ge 0.8$) to show complete metallic behavior at low temperatures. 

\begin{figure}
\includegraphics[width=80mm]{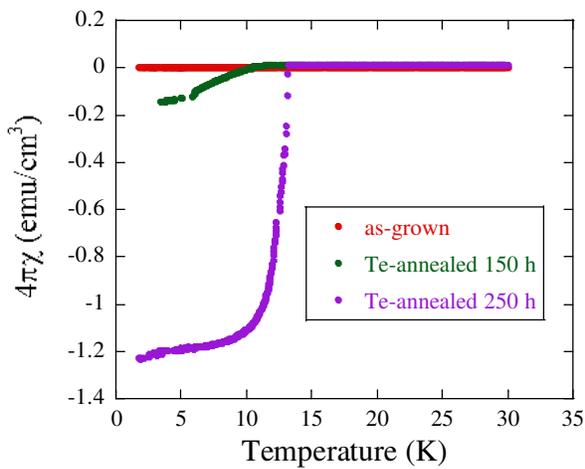}
\caption{\label{f5} (Color online) Temperature dependence of the magnetic susceptibility, $\chi$, of as-grown and annealed Fe$_{1.03}$Te$_{0.8}$Se$_{0.2}$ (nominal composition) single crystals. The measurements have been carried out with the zero-field-cooling procedure in a magnetic field of 100 Oe. perpendicular to the c-axis.}
\end{figure}

Figure \ref{f5} shows the temperature dependence of the magnetic susceptibility, $\chi$, measured with the zero-field-cooling procedure for as-grown and annealed Fe$_{1.03}$Te$_{0.8}$Se$_{0.2}$ (nominal composition) single crystals. The as-grown and vacuum-annealed samples (not shown) do not exhibit bulk superconductivity, although a very slight diamagnetic signal is observed below 6 K and 10 K, respectively. The superconductivity observed in $\rho_{ab}$ measurements for the vacuum-annealed sample may be filamentary. In contrast, when the as-grown sample is Te-annealed for 150 h, it shows bulk superconductivity below 11 K. The transition is, however, broad and the superconducting volume fraction is estimated as large as 15 $\%$. The sample that is Te-annealed for 250 h shows sharp superconducting transition at 13 K and a full magnetic shielding effect at lower temperatures. A slight difference in the ideal diamagnetism observed in this sample is due to the demagnetization effect. These results reveal that Te-annealing remarkably improves the superconductivity of Fe-based superconductors Fe$_{1+y}$Te$_{x}$Se$_{1-x}$.
  
\begin{table}
\caption{Chemical compositions, superconducting transition temperatures, $T_{c}$, and magnetic shielding volume fractions before and after annealing of the Fe$_{1.03}$Te$_{0.8}$Se$_{0.2}$ (nominal composition) single crystals. The chemical compositions are determined by EPMA measurements. $T_{c}$ is defined as the temperature showing zero resistivity.}
\label{t1}
\begin{center}
\begin{tabular}{cccc}
\hline
\multicolumn{1}{c}{Annealing} & \multicolumn{1}{c}{Fe:Te:Se($\pm0.02$)}& \multicolumn{1}{c}{$T_{c}$ (K)}& \multicolumn{1}{c}{Shielding volume fraction ($\%$)} \\
\hline
as-grown & 1.09:0.84:0.16 & --- & --- \\
vacuum-annealed 150 h & 1.09:0.83:0.17 & 10 & $\leq$ 1 \\
Te-annealed 150 h & 1.06:0.83:0.17 & 11 & 15 \\
Te-annealed 250 h & 1.00:0.82:0.18 & 13 & 100 \\
\hline
\end{tabular}
\end{center}
\end{table}

In order to investigate the origin, EPMA measurements have been carried out. Table \ref{t1} summarizes the EPMA results, $T_{c}$, and the magnetic shielding volume fractions before and after annealing of the Fe$_{1.03}$Te$_{0.8}$Se$_{0.2}$ (nominal composition) single crystals. The composition of the as-grown sample is estimated to be Fe$_{1.09}$Te$_{0.84}$Se$_{0.16}$, which is close to the nominal value but slightly Te-rich. The previous studies\cite{fang,kawasaki} on the c-axis lengths of Fe$_{1+y}$Te$_{x}$Se$_{1-x}$ show that it is $\approx$ 6.17 \AA \  for $x = 0.8$ and increases as $x$ increases. The c-axis length obtained for our as-grown sample is 6.20 \AA \ (Fig. \ref{f3}), which is slightly larger than 6.17 \AA. This is in accord with the observation that our as-grown sample is slightly Te-rich. The vacuum-annealing does not change the composition of the as-grown sample within the confines of experimental error. On the other hand, as can be seen in Fig. \ref{f3}, the scattering of the data for the c-axis length of the vacuum-annealed sample is decreased compared with that of the as-grown sample. These results may correlate with the indication that vacuum-annealing makes the crystal lattice more homogeneous\cite{noji2,taen}. As for the superconductivity, vacuum-annealing improves it, but the effect is not so great (Fig. \ref{f4}).  

\begin{figure}
\includegraphics[width=100mm]{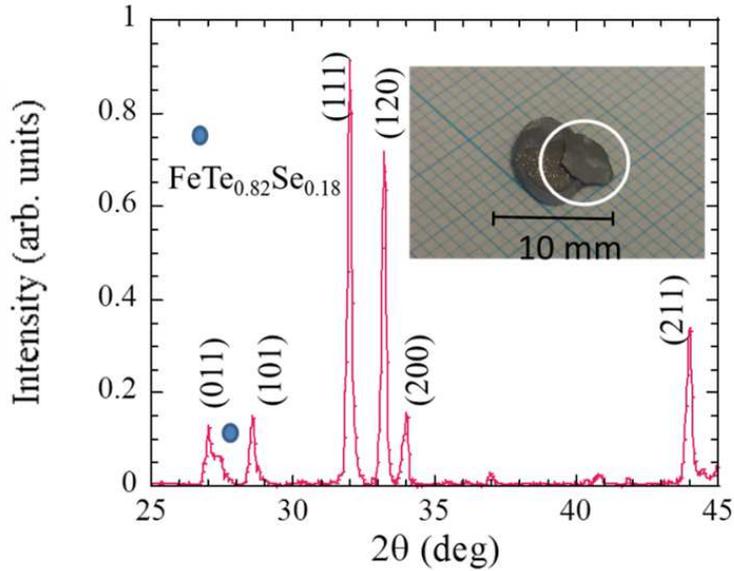}
\caption{\label{f6} (Color online) The XRD pattern of a surface portion of a Te-annealed sample. All peaks except the marked peak are assigned to those of FeTe$_2$. The marked peak indicates a $(101)$ peak of a mother FeTe$_{0.82}$Se$_{0.18}$ phase. The inset shows a photograph of the Te-annealed sample. The cleaved surface portion examined is surrounded by a circle.}
\end{figure}

In contrast, it is observed that Te-annealing obviously decreases the amount of excess Fe, $y$, from 0.09 to zero (Table \ref{t1}). It is also noticed that Te concentration, $x$, tends to decrease (from 0.84 to 0.82) along with Te-annealing (Table \ref{t1}). The decrease in $x$ interprets the decrease in c-axis length (Fig. \ref{f3}). Then, why is excess Fe decreased by Te-annealing? After Te-annealing, the surface of the crystal is observed to be grayish as shown in the inset of Fig. \ref{f6}. This indicates that some type of Te-compound is formed on the surface of the crystal. The surface portion has been carefully cleaved out for XRD measurements as shown in the inset of Fig. \ref{f6}. Figure \ref{f6} shows a powder XRD pattern of the specimen. Almost all peaks are assigned to those of FeTe$_2$\cite{yama}. Thus, it has been found that new layers of FeTe$_2$ are formed on the sample surface. The marked peak is assigned to a $(101)$ peak of a mother FeTe$_{0.82}$Se$_{0.18}$ crystal. The result that the surface portion is composed of FeTe$_2$ is also confirmed by the EPMA measurements. During Te-annealing, the excess Fe atoms in the as-grown crystal may take part in forming FeTe$_2$ with surrounding Te atoms at the crystal surface. After Te-annealing, the surface portion has been cleaved away from the crystals. The decrease in excess Fe concentration has been caused by this process. It has been argued that the excess Fe atoms disturb the short-range magnetic order in the adjacent Fe\textit{X} (\textit{X}:Te or Se) layers, which results in charge-carrier localization and suppression of superconductivity\cite{liu}. The observation of the metallic resistivity for our Te-annealed sample is consistent with this scenario. Therefore, we conclude that the remarkable improvement in superconductivity by Te-annealing is due to the removal of the excess Fe atoms.  


In summary, single crystals of Fe$_{1.03}$Te$_{0.8}$Se$_{0.2}$ were grown using the Bridgman method and then they were annealed under Te vapor at 400 $^\circ$C for 250 h. Their resistivity and magnetic susceptibility measurements show that they are metallic in the normal state and bulk superconducting below 13 K, respectively, whereas the as-grown sample is not. Our EPMA analysis show that the excess Fe that existed in the as-grown sample was completely removed by the annealing. It should be emphasized that this new annealing method is highly effective in producing superconductivity and also applicable to large bulk single crystals. 
\begin{acknowledgment}

We thank T. Noji and  Y. Koike, Tohoku University, and H. J. Im, Hirosaki University, for helpful discussions. We thank S. Kobayashi, Iwate University, for his help in magnetization measurements. We thank S. Ando, Hirosaki University, for his help in the EPMA measurements.

\end{acknowledgment}

\end{document}